\documentclass[lettersize,journal]{IEEEtran}
\usepackage{amsmath,amsfonts}
\usepackage{algorithmic}
\usepackage{algorithm}
\usepackage{array}
\usepackage[caption=false,font=normalsize,labelfont=sf,textfont=sf]{subfig}
\usepackage{textcomp}
\usepackage{stfloats}
\usepackage{url}
\usepackage{verbatim}

\usepackage{cite}
\usepackage{graphicx}

\usepackage{pgfplots}
\usepackage{tikz}
\usepackage{csvsimple}
\usepackage{multirow}

\def\BibTeX{{\rm B\kern-.05em{\sc i\kern-.025em b}\kern-.08em
    T\kern-.1667em\lower.7ex\hbox{E}\kern-.125emX}}
\usepackage{balance}

\begin{document}

\title{Predicting Stock Price Movement as an Image Classification Problem}

\author{Matej Steinbacher
\thanks{M. Steinbacher is an independent researcher and is affiliated with the Pixlifai. He can be contacted at matej.steinbacher@gmail.com.}
}



\maketitle

\begin{abstract}
  The paper studies intraday price movement of stocks that is considered as an image classification problem. Using a CNN-based model we make a compelling case for the high-level relationship between the first hour of trading and the close. The algorithm managed to adequately separate between the two opposing classes and investing according to the algorithm’s predictions outperformed all alternative constructs but the theoretical maximum. To support the thesis, we ran several additional tests. The findings in the paper highlight the suitability of computer vision techniques for studying financial markets and in particular prediction of stock price movements.
\end{abstract}

\begin{IEEEkeywords}
  Deep convolutional neural networks, MobileNet-V2, image classification, stock prices, investment problem.
\end{IEEEkeywords}

\section{Introduction}
\IEEEPARstart{G}{aining} an ability to predict price movement of stocks and other tradables remains an imperative for an academic and professional community. Markets are volatile and inherently subject to various economic, political, psychological factors which makes such predictions a challenging task, especially at the intraday level. For instance, on September 21, 2022 at the highly anticipated talk of the Federal Reserve’s Chair Jerome Powell about the interest rates hike, the NASDAQ index initially fell by almost 1.5\% within the first 30-minute time frame, then within the next 30 minutes grew more than 2\% to fall back by almost 1.7\% within the additional 30 minute-period and to close the day at the day’s low by falling for an additional 1.3\% \cite{SubinReinicke_2022}.

To address the question many different financial and economic variables have been proposed as potential predictors, like valuation ratios such as the dividend-price ratio, dividend yield, earnings-price ratio, book-to-market ratio, various interest rates and interest rate spreads, macroeconomic variables including inflation and industrial production. See, for instance, \cite{bodie1976common, fama1977asset, campbell1987stock, keim1986predicting, french1987expected, barberis2000investing, stambaugh1999predictive, welch2008comprehensive}. \cite{barberis2003survey} is a nice summary of the role of psychology in asset pricing. These papers contributed to the fundamental understanding of financial markets that at most explained general relationships between variables but turned out to be of limited use for prediction of price movements of individual stocks.

Lately, general advancements within the field of artificial intelligence and in particular of computer vision entered the field of financial markets as well. Some of the examples include \cite{eapen2019novel} that combined pipelines of 1-dimensional CNN model with bi-directional LSTM units. \cite{tsantekidis2017forecasting} used a CNN model on a limit order book, i.e., a list of buy and sell orders, to predict the direction of price change. \cite{sim2019} used LeNet-5 model for predicting a boolean direction of stock-price’s change using the S\&P500’s index data. Authors used closing prices on a minute basis and calculated nine technical indicators as input variables whose data for 30 minutes was plotted as images. Similarly, \cite{gudelek2017deep} calculated 28 technical indicators over 28 days to get a $28\times28$ square matrix that was processed with a CNN-based model or \cite{sezer2018algorithmic} who used $15\times15$ frame. \cite{long2019deep} integrated elements of CNN and RNN neurons into a multi-filter structure to analyze Chinese stock market index CSI300 while \cite{chen2021novel} proposed a graph convolutional CNN network that let them combine the general market information and the individual stock information into a joint feature. For an extensive overview, see \cite{hu2021survey, sezer2020financial}.

In the paper we study the predictability of intraday stock returns based on the data of the largest NASDAQ stocks. A decision for trading a stock is considered as an image-classification problem with all intraday yields classified in three classes that indicate if a stock is a buy, a sell or too volatile to call. The model is built on the top of the MobileNet-V2 \cite{sandler2018mobilenetv2}. To feed the model, all financial data was visualized in the form of an image. Our construction of the problem derives from an assumption about an existence of a stable, high-level relationship between the first hour of trading and the closing price.

Image classification is a standard computer vision domain. Its methods derive from an assumption that processing of a large set of images can extract some high-level abstraction from the data, add to the understanding of the problem at stake and (hopefully) facilitate the decision-making. CNN-based models, a topic of the paper, have three main advantages over traditional neural networks: parameter sharing, sparse interactions and equivalent representations \cite{goodfellow2016deep}. To fully utilize the two-dimensional structure of an input data, local connections and shared weights in the network are utilized, instead of traditional fully connected networks.

\IEEEpubidadjcol Simulation results make a compelling case in favor of our approach since investing according to the algorithm’s predictions outperformed all alternative architectures and was short only to the algorithm’s theoretical maximum. General accuracy score of the testing was 51.5\% for the sample of three classes while precision score of the two most important classes C1 and C2 equalled 53.3\% and 58.3\%, respectively, when corrected for the size. The algorithm managed to separate classes on opposing poles very well with a vast majority of their false classifications being classified into the third and neutral class C0 that was added as a “buffer zone” between the two. Adding stricter conditions for classification improved scores even further. For instance, requiring at least a 95\% “approval” score to make the call led to 88.7\% accuracy rate. However, it classified only 71 observations or 2.2\% of the original dataset. A clear indication of trading a quantity for quality. Altogether, results make a strong evidence in favor of modeling financial markets with deep image-based models the way it was done in the paper. Results also revealed a need for rethinking the use of performance metrics that are traditionally used in modeling deep neural networks that seem insufficient.\IEEEpubidadjcol

The paper proceeds as follows. Chapter 2 describes the data. The model was built using the data of the largest NASDAQ-listed stocks. Chapter 3 outlines the model. We used the MobileNet-V2 of Google as the base. Extensive testing and simulations are done in Chapter 4. Results of the real-time testing are shown and discussed in Chapter 5. Last chapter concludes.

\section{The Data}
\noindent The initial dataset consists of 1000 largest NASDAQ listings measured by market capitalization. The data was collected between January 14 and July 1, 2022 using Python’s open source library yfinance \cite{yfinance_2022} and was visualized with Python’s open source library mplfinance \cite{mplfinance_2022}. Regular trading with stocks is open Monday through Friday between 9:30 a.m. to 4:00 p.m. ET. The data was collected on a 5-minute basis and for each time-interval an information for open, close, low, high and the volume was used. Any incomplete unit of data was dropped.
  
\begin{figure}[!t]
    \centering
    \includegraphics[width=2.5in]{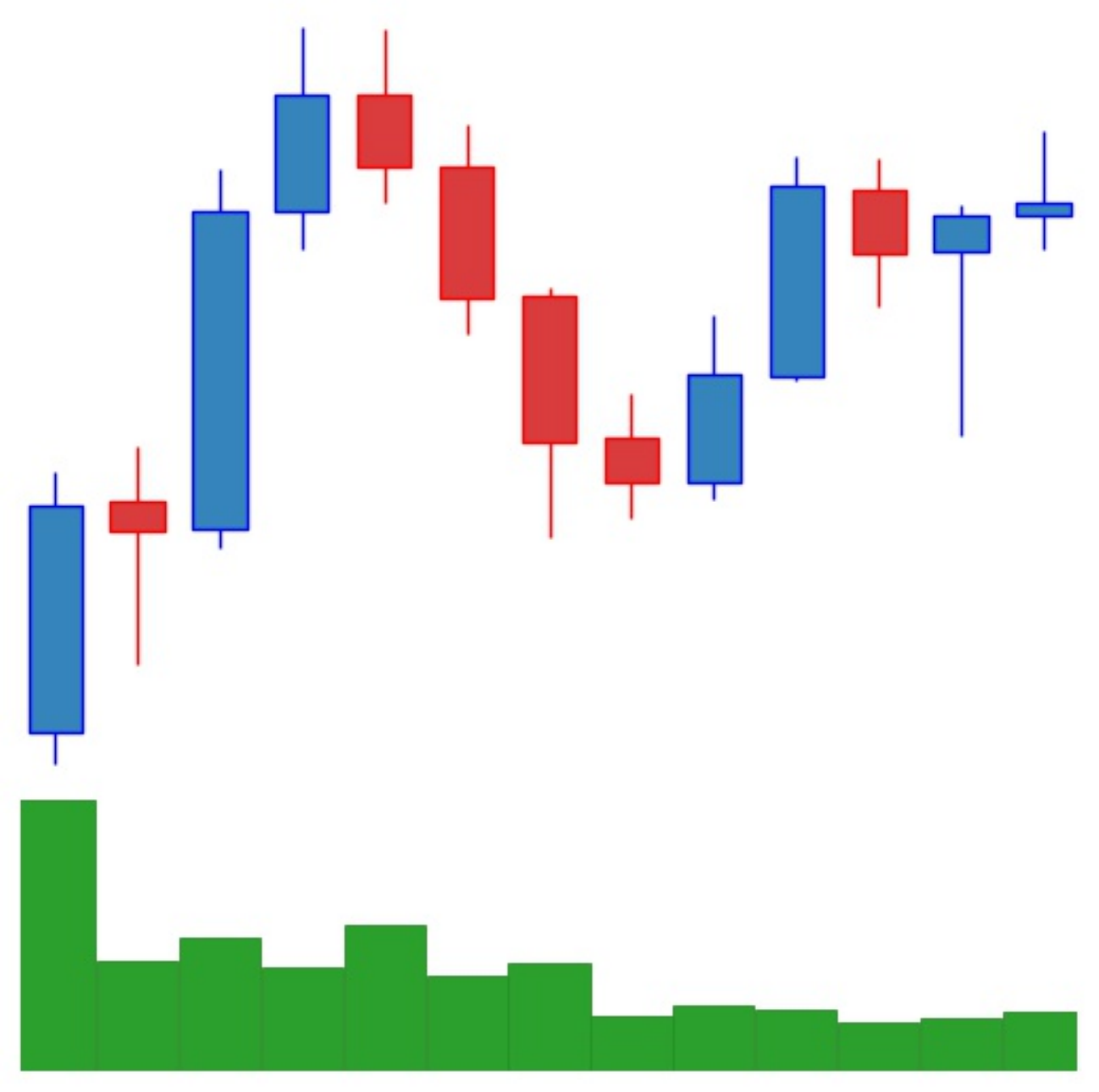}
    \caption{A Sample Image}
    \label{fig:sample_image}
\end{figure}

Each unit of data $s_i = \{x_i,y_i\}$ is defined as a pair of an input image $x_i$ and a corresponding label $y_i$. The sample of the input image is shown in figure \ref{fig:sample_image}.

The image shows a sequence of candlesticks and corresponding volume bars for each 5-minute interval from the first hour of regular trading. A candlestick is a type of a price chart that is used in technical analysis and displays prices of an asset for the high, low, open, and closing during a specified interval of time. Candlesticks are often used by traders in a search for price patterns. Volume is plotted in a form of a bar and shows a number of shares of an asset that was traded within a specified period of time.

Units were labeled according to the following relations:

\begin{equation*}
    y_{i} =
    \begin{cases}
    1,&{\text{if}}\ \frac{p_T^i}{p^i_0} \geq 1.02,\\
    2,&{\text{if}}\ \frac{p_T^i}{p^i_0} \leq 0.98,\\
    {0,}&{\text{otherwise.}} 
    \end{cases}
\end{equation*}
    
where $y_i$ is a label of an observation $i$, $p^i_0$ stands for the close price of an asset $i$ right after the first hour of regular trading and $p^i_T$ is asset $i$’s close price at the close of regular trading hours.

\begin{table*}[!htp]
    \begin{center}

  \caption{Datasets Summary Statistics}
  \label{Table: Datasets: Summary Statistics}
  \begin{tabular}{|l|c|c|c|c|c|c|c|c|c|c|c|c|}
    \hline
    \multicolumn{1}{|c|}{} & \multicolumn{4}{c|}{$S_l$} & \multicolumn{4}{c|}{$S_v$} & \multicolumn{4}{c|}{$S_t$}\\
    \hline
    \multicolumn{1}{|c|}{} & \textrm{ALL} & \textrm{0} & \textrm{1} & \textrm{2} & \textrm{ALL} & \textrm{0} & \textrm{1} & \textrm{2} & \textrm{ALL} & \textrm{0} & \textrm{1} & \textrm{2} \\
    \hline AVG & 1.000 & 1.000 & 1.033 & 0.968 & 1.000 & 1.000 & 1.033 & 0.968 & 1.003 & 1.000 & 1.033 & 0.969\\
    \hline MEDIAN & 0.999 & 1.000 & 1.028 & 0.972 & 1.000 & 1.000 & 1.028 & 0.972 & 1.002 & 1.000 & 1.028 & 0.973\\
    \hline SD & 0.029 & 0.010 & 0.014 & 0.013 & 0.029 & 0.009 & 0.014 & 0.012 & 0.020 & 0.010 & 0.015 & 0.013\\
    \hline MIN & 0.745 & 0.980 & 1.020 & 0.745 & 0.875 & 0.980 & 1.020 & 0.875 & 0.876 & 0.980 & 1.020 & 0.876\\
    \hline MAX & 1.205 & 1.020 & 1.205 & 0.980 & 1.127 & 1.020 & 1.127 & 0.980 & 1.152 & 1.020 & 1.152 & 0.980\\
    \hline Q1 & 0.977 & 0.993 & 1.024 & 0.964 & 0.977 & 0.993 & 1.024 & 0.964 & 0.992 & 0.993 & 1.024 & 0.966\\
    \hline Q3 & 1.023 & 1.007 & 1.036 & 0.977 & 1.023 & 1.007 & 1.037 & 0.977 & 1.013 & 1.007 & 1.038 & 0.977\\
    \hline N & 14,175 & 5,000 & 4,457 & 4,718 & 3,983 & 1,400 & 1,255 & 1,328 & 3,179 & 2,314 & 566 & 299\\
    \hline
  \end{tabular}
\end{center}
\end{table*}

The dataset was split into three non-intersecting subsamples that were later used for training, validation and testing. To control for the class imbalance, subsamples were downsampled by removing random units. Table \ref{Table: Datasets: Summary Statistics} shows summary statistics of subsamples.

\section{The Model}
\noindent We used the MobileNet-V2 (MNET-2) \cite{sandler2018mobilenetv2}. This is a small CNN model which uses a combination of depthwise separable convolutions and inverted residual blocks where the shortcut connections are between the thin bottleneck layers. The whole architecture keeps the size of a tensor relatively small, though increasing in channels but tiny in spatial dimensions, mostly due to bottleneck layers that keep connections between the blocks in low dimension. The model builds on a premise that “manifolds of interest” can be embedded in low-dimensional subspaces and was chosen for three main reasons: as a lighter model with smaller amount of parameters it should contain the potential presence of overfitting; due to smaller computational costs and performance.

The model was trained on the top of the last linear layer of the original model with a dropout regularization rate of 20\%. Final, fully connected layer has no nonlinearity and feeds directly into the softmax layer for classification. The softmax function maps a vector of logits to a posterior probability distribution. The model was optimized with the stochastic gradient descent (SGD), with a learning rate of 0.001 that decayed by the factor of 10 for every 30 epochs, a momentum of 0.9 and a weight decay of 0.0001 with Nesterov momentum.

Input images have a fixed resolution of $224 \times 224$ and 3 color channels as required by the model with each pixel defined on the $[0, 255]$ domain. The dataset had been normalized with $x'=\frac{(x/255)-\mu}{\sigma}$ where $\mu = [0.485,0.456,0.406]$ and $\sigma =[0.229,0.224,0.225]$.

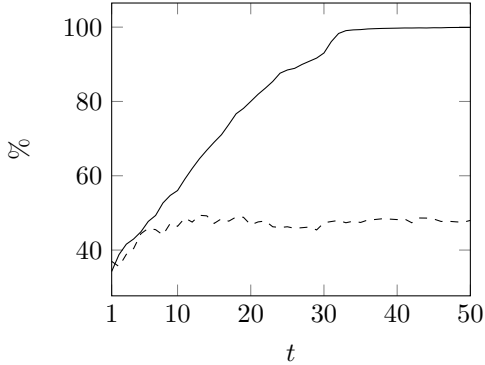
\begin{figure}[!htp]
    \pgfplotstableread[col sep=comma,]{res/training_results.csv}\datatable
    \centering
    \begin{tikzpicture}
        \begin{axis}[
            xmin=1, xmax=50,
            width=2.5in,
            xlabel=$t$,
            ylabel=$\%$,
            xtick={1,10,20,...,50},scaled x ticks=false,
            ]
            \addplot [
                color=black,
                mark size=1pt
                ] table[x=N, y=Training]{\datatable};
            \addplot [
                dashed,
                color=black,
                mark size=1pt
                ] table[x=N, y=Validation]{\datatable};
        \end{axis}
    \end{tikzpicture}
    \caption{Training Results}
    \label{train_val_results}
\end{figure}

The whole modeling was done in the PyTorch and was run on a single Nvidia’s GeForce 930M GPU with 2GB of memory. Each model was trained for $T=50$ epochs. Results are shown on the Fig. \ref{train_val_results}.

The Fig. \ref{train_val_results} shows accuracy scores during the training (solid line) and validation (dashed line) phases. The top performing model achieved 49.36\% accuracy at the validation level. Given three balanced classes of the input data, the result outperforms random guessing by 48\% at the aggregate level.

The shape of both curves indicate that the training was a subject of overfitting. An overfitted model may fail to properly generalize features that it is supposed to learn and instead fits the idiosyncrasies of the training sample itself. Such model would perform well during the training but unsatisfactorily to any other data but the one on which it was trained. Overfitting is a general issue in the domain of supervised machine learning and cannot be avoided\cite{pmlr-v119-rice20a}.

To contain the effect of overfitting, we added a dropout layer that discarded 20\% of random units to the end of the network just prior to the classification. Dropout is generally considered very effective technique against overfitting\cite{JMLR:v15:srivastava14a}, \cite{warde2013}.

\section{Testing and Simulations}
\subsection{Testing}
\noindent Results of the testing are shown in Tables \ref{tbl:performance_results} and \ref{Table:performance_metrics}. The algorithm managed to properly classify 1636 out of 3179 observations for an overall accuracy of 51.46\%. The algorithm left the largest class C0 with a 72.8\% share of observations underrepresented by classifying to it "only" 47.7\% of all predictions. On the other side, C1 gained its relative importance against both other classes. The algorithm managed to separate between the opposite classes C1 and C2 fairly well and made most of their false classifications into the neutral class C0. For instance, only 18.7\% of C2 was misclassified into C1, while 43.8\% into C0. Also, only 10\% of C1 observations were classified as C2, while 32.7\% as C0.

\begin{table}[!ht]
    \centering
    \caption{Confusion Matrix}
    \label{tbl:performance_results}
    \begin{tabular}{|c|c|c|c|c|}
        \hline True / Prediction  & 0 & 1 & 2 & SUM\\
        \hline 0 & 1200 & 728 & 386 & 2314\\
        \hline 1 & 185 & 324 & 57 & 566\\
        \hline 2 & 131 & 56 & 112 & 299\\
        \hline SUM & 1516 & 1108 & 555 & 3179\\
        \hline
    \end{tabular}
\end{table}

\begin{table}[!ht]
    \centering
    \caption{Performance Metrics}
    \label{Table:performance_metrics}
    \begin{tabular}{|c|c|c|c|c|}
        \hline & precision & recall & f1-score & support\\
        \hline 0 & 0.79 & 0.52 & 0.63 & 2314\\
        \hline 1 & 0.29 & 0.57 & 0.39 & 566\\
        \hline 2 & 0.20 & 0.37 & 0.26 & 299\\
        \hline
    \end{tabular}
\end{table}

Although the general accuracy level is the most intuitive performance measure, it is less relevant for us. It is so mainly for two reasons: (1) unbalanced dataset, (2) unequal interest for classes and their predictions. First, if all observations were classified into the C0, the algorithm would boast with a 72.8\% accuracy. Second, a false prediction of C0 as a C1 is undesirable but less troubling because C0 is distributed fairly symmetrically with a 0\% expected yield. On the other hand, a false prediction of C2 as C1 would result in an unavoidable loss of at least 2.5\% and 3.1\% on average.

To calculate the average yield, we combine results from the Table~\ref{tbl:performance_results} with the data from Table \ref{Table: Datasets: Summary Statistics}. It equals $E(\gamma | \pi, \overline\gamma) = \sum_{k=1}^J \pi_{k,1} \overline{\gamma}_k = 0.8\%$ with $\overline{\gamma}=[\overline{\gamma}_1 \ \dots \ \overline{\gamma}_J]$ average yield of class $k$ and $\pi_{i,1} = [0.657 \ 0.292 \ 0.051]$. This value is larger than the average value of the dataset that stands at $0.3\%$.

\subsection{Macro Analysis}
\noindent Assume that each class $C_j$ is distributed according to $F_j$ and let $\pi_{i,j}$ be a probability that a yield of class $i$ is classified into the class $j$ by the algorithm with $\pi_{i,j} \in [0,1]$ and $\sum_{j=1}^J \pi_{i,j} = 1 \forall i$. Assume further that each class $j$ has its own propensity to invest $\alpha_j \in [0,1]$. Then the expected yield of a random draw from $X$ can be written as

\begin{equation*}
  E(\gamma_i|x_i \in X)= \sum_{j=1}^J \left( \frac{|C_j|}{\sum_{k=1}^J|C_k|}E(X | F_j) \cdot \pi_{j,k} \cdot \alpha_{k} \right),
\end{equation*}

where $k \in \{1,...,J\}$ and $|C_j|$ is a size of the class $j$. We can show that there exists a size of classes that result in a positive expected yield. For instance, assume 3 classes with the following expected yields: $\gamma_1 \geq 0$ and $\gamma_3 \leq 0$, while $\gamma_2$ can take any value. $n_i=\frac{|N_i|}{\sum_{k=1}^3|N_k|}$ is a proportion of class $i$ observations and let only a classification to $C_1$ get invested or $\alpha_1=1$, $\alpha_2=0$ and $\alpha_3=0$ propensities to invest, then we can write

\begin{equation*}
    n_1 \cdot \gamma_1 \cdot \pi_1 + n_2 \cdot \gamma_2 \cdot \pi_2 + n_3 \cdot \gamma_3 \cdot \pi_3 \geq 0,
\end{equation*}

subject to $n_1+n_2+n_3=1$. $\pi_i$ indicates a probability that class $i$ is classified as class 1.

Then $n_1 \geq - \frac{\gamma_2 \cdot \pi_2}{\gamma_1 \cdot \pi_1} \cdot n_2 - \frac{\gamma_3 \cdot \pi_3}{\gamma_1 \cdot \pi_1} \cdot n_3$ which can be rewritten to

\begin{equation*}
    \frac{n_1}{n_2} \geq - k_2 - k_3 \cdot \frac{n_3}{n_2}.    
\end{equation*}
  
$k_2$ would usually be very small, close to 0, while $k_3 \leq0$. This means that the size of the positively yielded class $n_1$ should be proportional to the size of the negative $n_3$. For instance, if the expected yield of the positive $\gamma_1=0.03$ and the negative $\gamma_3=-0.03$, then, ceteris paribus, $n_1$ would need to be at least approximately one-third of the $n_3$ to break even on average.

To test the written, we ran three batches of Monte Carlo simulations with 10,000 repetitions each. $X_j$ for class $j=1,...,J$ followed a truncated normal distribution $X_j \sim N(\mu_j,\sigma_j^2; a_j,b_j)$ with mean $\mu_j$, variance $\sigma_j^2$ and $- \infty \leq a_j < b_j \leq \infty$. Parameters used during the simulation are given in Table \ref{table:monte_carlo_parameters}.

\begin{table}[h]
  \caption{Monte Carlo Parameters}
  \label{table:monte_carlo_parameters}
  \centering
  \begin{tabular}{|c|c|c|c|c|c|c|c|}
    \hline Class & $\mu$ & $\sigma$ & a & b & Exp1 & Exp2 & Exp3\\
    \hline 1 & 0.03 & 0.015 & 0.02 & 0.15 & 33 & 100 & 50\\
    \hline 2 & 0.0 & 0.01 & -0.02 & 0.02 & 10 & 10 & 10\\
    \hline 3 & -0.03 & 0.015 & -0.15 & -0.02 & 100 & 100 & 300\\
    \hline
  \end{tabular}
\end{table}
  
\begin{figure}[!htp]
    \centering
    \pgfplotstableread[col sep=comma,]{res/monte_carlo/exp1.csv}\datatable
    \begin{tikzpicture}
        \begin{axis}[
            xmin=472, xmax=1732, ymin=1, ymax=60,
            width=2.5in,
            xlabel=$v_T$,
            ylabel=$N$
            ]
            \addplot [
                only marks,
                color=black,
                mark size=0.5pt
                ] table[x=exp1, y=N]{\datatable};
        \end{axis}
    \end{tikzpicture}
    \caption{Monte Carlo Simulations: Exp1}
    \label{fig:monte_carlo_exp1}
\end{figure}
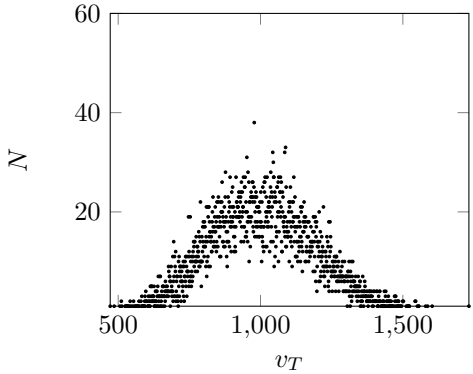

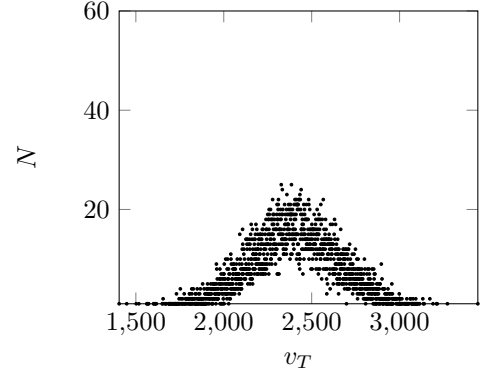
\begin{figure}[!htp]
    \centering
    \pgfplotstableread[col sep=comma,]{res/monte_carlo/exp2.csv}\datatable
    \begin{tikzpicture}
        \begin{axis}[
            xmin=1405, xmax=3445, ymin=1, ymax=60,
            width=2.5in,
            xlabel=$v_T$,
            ylabel=$N$
            ]
            \addplot [
                only marks,
                color=black,
                mark size=0.5pt
                ] table[x=exp2, y=N]{\datatable};
        \end{axis}
    \end{tikzpicture}
    \caption{Monte Carlo Simulations: Exp2}
    \label{fig:monte_carlo_exp2}
\end{figure}

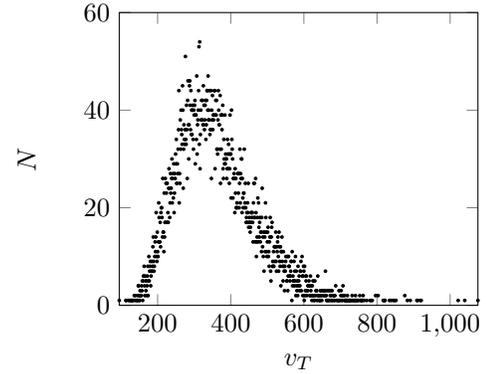
\begin{figure}[!htp]
    \centering
    \pgfplotstableread[col sep=comma,]{res/monte_carlo/exp3.csv}\datatable
    \begin{tikzpicture}
        \begin{axis}[
            xmin=95, xmax=1076, ymin=0, ymax=60,
            width=2.5in,
            xlabel=$v_T$,
            ylabel=$N$
            ]
            \addplot [
                only marks,
                color=black,
                mark size=0.5pt
                ] table[x=exp3, y=N]{\datatable};
        \end{axis}
    \end{tikzpicture}
    \caption{Monte Carlo Simulations: Exp3}
    \label{fig:monte_carlo_exp3}
\end{figure}

Initial investment in all cases was 1000USD and all subsequent investments equalled $\min(v_t,1000)$ given the corresponding probabilities $\pi_{i,1} = [0.572, 0.315, 0.187]$ that were calculated from the Table \ref{tbl:performance_results}. $v_t$ is the whole accumulated sum up to time $t$. Simulation results are given in Figs. \ref{fig:monte_carlo_exp1}, \ref{fig:monte_carlo_exp2}, \ref{fig:monte_carlo_exp3}.

Fig. \ref{fig:monte_carlo_exp1} shows simulation results when the structure of all classes complied with the closed-form mathematical solution. That is, to break-even the size of positively-yielded class 1 should be one-third of the size of the negatively-yielded class 3, ceteris paribus. $E(x \in C_2)=0$ does not affect the result and was, hence, kept fixed. Running the Monte Carlo turned initial 1000USD to 1003USD on average in 143 draws for a minor yield of $E(x \in X| \text{Exp1}) =0.002\%$ per draw. The result is basically in line with the theoretical solution.

Fig. \ref{fig:monte_carlo_exp2} shows simulation results with equal sizes of positively and negatively yielded classes, ceteris paribus. The run turned initial investment into 2400USD on average. The result is a clear consequence of different buy probabilities for both classes that is heavily inclined towards the true buy class. Finally, in Fig. \ref{fig:monte_carlo_exp3}, a very bearish daily sentiment, the run resulted in a heavy loss that turned initial 1000USD to 361USD on average. A clear implication of the latter is that there also exists a state when trading with the algorithm would lead to a cumulative loss on average and when it would be better to altogether drop trading, ceteris paribus.

\subsection{Micro Analysis}
\noindent In the previous section all values were artificially generated from accompanying distributions. Herein, all data will be taken from the testing dataset as it appeared in real trading.

\begin{table}[h]
  \caption{Algorithm's Predictions: Statistics}
  \label{Table: Algorithm's Predictions}
  \centering
  \begin{tabular}{|c|c|c|c|}
    \hline
    \multicolumn{1}{|c|}{} & \multicolumn{1}{c|}{C0} & \multicolumn{1}{c|}{C1} & \multicolumn{1}{c|}{C2}\\
    \hline AVG & 1.0011 & 1.0095 & 0.9955\\
    \hline MEDIAN & 1.0000 & 1.0070 & 0.9964\\
    \hline SD & 0.0171 & 0.0216 & 0.0209\\
    \hline MIN & 0.9040 & 0.9330 & 0.8756\\
    \hline MAX & 1.1080 & 1.1520 & 1.1217\\
    \hline Q1 & 0.9920 & 0.9958 & 0.9829\\
    \hline Q3 & 1.0100 & 1.0230 & 1.0064\\
    \hline N & 1516 & 1108 & 555\\
  \hline
  \end{tabular}
\end{table}

Table \ref{Table: Algorithm's Predictions} summarizes the performance of the algorithm during the testing. Out of 3179 observations, 1516 were classified as C0 with an average gain per element $0.1\%$, 1108 as C1 with an average gain per element of $0.95\%$ and 555 as C2 with an average loss per element of $-0.45\%$.

Considering a small gain for C0 as a first-order approximation of the zero yield, then the calculated directions are in line with expectations, which means that the algorithm managed to separate between classes as a whole. However, it heavily compressed the magnitudes. For instance, the average gain of a predicted C1 stands at no more than 28.8\% of the true C1.

Only predictions of C1 initiate an activity. Altogether, 1108 C1 predictions were made, of which 722 with a non-negative yield and 386 with a negative yield. A negative yield can be a consequence of a false C0 or C2 prediction. The average yield of the C1 prediction is just shy of 1\% per trade. The value is higher than the average yield at the “macro” level that was just calculated at 0.8\% per trade and significantly higher than the average yield of the testing dataset of 0.3\% (Table \ref{Table: Datasets: Summary Statistics}).

\subsection{Distribution of Predictions}
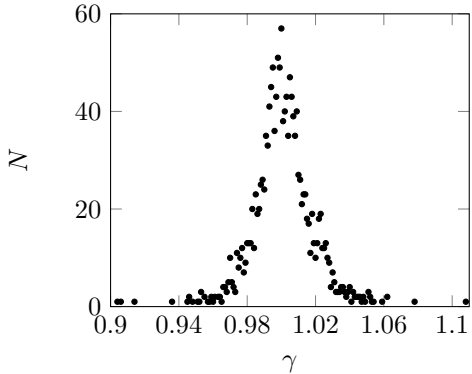
\begin{figure}[!htp]
    \centering
    \pgfplotstableread[col sep=comma,]{res/dist_of_predictions/c0.csv}\datatable
    \begin{tikzpicture}
        \begin{axis}[
            xmin=0.9, xmax=1.11, ymin=0, ymax=60,
            width=2.5in,
            xlabel=$\gamma$,
            ylabel=$N$,
            xtick={0.9,0.94,...,1.11},scaled x ticks=false,
            ]
            \addplot [
                color=black,
                mark size=1pt,
                only marks
                ] table[meta=count]{\datatable};
        \end{axis}
    \end{tikzpicture}
    \caption{Distributions of Class Predictions: C0}
    \label{fig:C0_dist_predictions}
\end{figure}

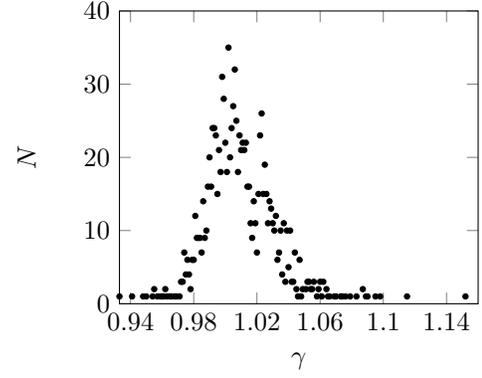
\begin{figure}[!htp]
    \centering
    \pgfplotstableread[col sep=comma,]{res/dist_of_predictions/c1.csv}\datatable
    \begin{tikzpicture}
        \begin{axis}[
            xmin=0.933, xmax=1.16, ymin=0, ymax=40,
            width=2.5in,
            xlabel=$\gamma$,
            ylabel=$N$,
            xtick={0.94,0.98,...,1.15},scaled x ticks=false,
            ]
            \addplot [
                color=black,
                mark size=1pt,
                only marks
                ] table[meta=count]{\datatable};
        \end{axis}
    \end{tikzpicture}
    \caption{Distributions of Class Predictions: C1}
    \label{fig:C1_dist_predictions}
\end{figure}

\begin{figure}[!htp]
    \centering
    \pgfplotstableread[col sep=comma,]{res/dist_of_predictions/c2.csv}\datatable
    \begin{tikzpicture}
        \begin{axis}[
            xmin=0.87, xmax=1.122, ymin=0, ymax=20,
            width=2.5in,
            xlabel=$\gamma$,
            ylabel=$N$,
            xtick={0.87,0.92,...,1.1},scaled x ticks=false,
            ]
            \addplot [
                color=black,
                mark size=1pt,
                only marks
                ] table[meta=count]{\datatable};
        \end{axis}
    \end{tikzpicture}
    \caption{Distributions of Class Predictions: C2}
    \label{fig:C2_dist_predictions}
\end{figure}
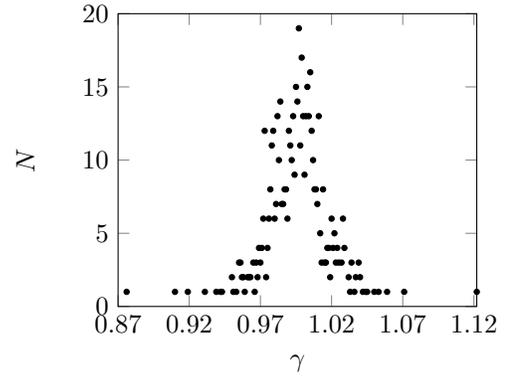

\noindent Distributions of predicted classes are shown on Figs. \ref{fig:C0_dist_predictions}, \ref{fig:C1_dist_predictions} and \ref{fig:C2_dist_predictions}. Frequencies were calculated on a discretized data with increments set at $0.001$ or $0.1\%$.

Shape of the C0 looks fairly symmetric and centered around $1.0$ with a skewness of $0.027$ but does not come from the normal distribution.{\footnote{Shapiro-Wilk $W = 0.966986$, with $p = 4.06126e-18$.}} C1 exhibits positive skewness $(0.845)$ with most values in the positive territory and C2 negative skewness $(-0.024)$ with the majority of the distribution in the territory of negative yields.

\subsection{Class Predictions and Yields}
\noindent Up to now class-prediction probabilities were considered exogenous at the level of particular class. In this section the assumption will be dropped. To be able to do the testing, we need to preprocess the data. First, to contain the presence of potential outliers, the data will be truncated as follows

\begin{equation*}
    {f(x|a=0.95,b=1.05)} =
    \begin{cases}
    a,&{\text{if}}\ x \leq a,\\ 
    b,&{\text{if}}\ x \geq b,\\ 
    {x,}&{\text{otherwise.}} 
    \end{cases}
\end{equation*}
    
\begin{figure}[!htp]
    \centering
    \pgfplotstableread[col sep=comma,]{res/class_predictions_yields.csv}\datatable
    \begin{tikzpicture}
        \begin{axis}[
            xmin=0.95, xmax=1.05, ymin=0, ymax=1,
            width=2.5in,
            xlabel=$\gamma$,
            ylabel=$\%$
            ]
            \addplot [
                only marks,
                color=black,
                mark size=1pt
                ] table[x=yield, y=C0]{\datatable};
        \end{axis}
    \end{tikzpicture}
    \caption{Proportions of Class Predictions per Yield: C0}
    \label{fig:C0_predictions}
\end{figure}
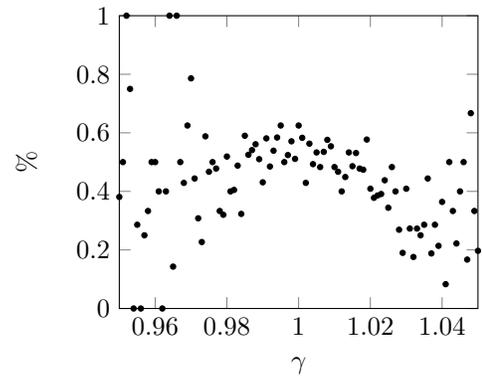

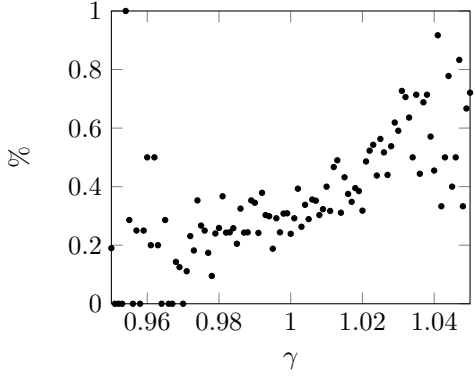
\begin{figure}[!htp]
    \centering
    \pgfplotstableread[col sep=comma,]{res/class_predictions_yields.csv}\datatable
    \begin{tikzpicture}
        \begin{axis}[
            xmin=0.95, xmax=1.05, ymin=0, ymax=1,
            width=2.5in,
            xlabel=$\gamma$,
            ylabel=$\%$
            ]
            \addplot [
                only marks,
                color=black,
                mark size=1pt
                ] table[x=yield, y=C1]{\datatable};
        \end{axis}
    \end{tikzpicture}
    \caption{Proportions of Class Predictions per Yield: C1}
    \label{fig:C1_predictions}
\end{figure}

\begin{figure}[!htp]
    \centering
    \pgfplotstableread[col sep=comma,]{res/class_predictions_yields.csv}\datatable
    \begin{tikzpicture}
        \begin{axis}[
            xmin=0.95, xmax=1.05, ymin=0, ymax=1,
            width=2.5in,
            xlabel=$\gamma$,
            ylabel=$\%$
            ]
            \addplot [
                only marks,
                color=black,
                mark size=1pt
                ] table[x=yield, y=C2]{\datatable};
        \end{axis}
    \end{tikzpicture}
    \caption{Proportions of Class Predictions per Yield: C2}
    \label{fig:C2_predictions}
\end{figure}
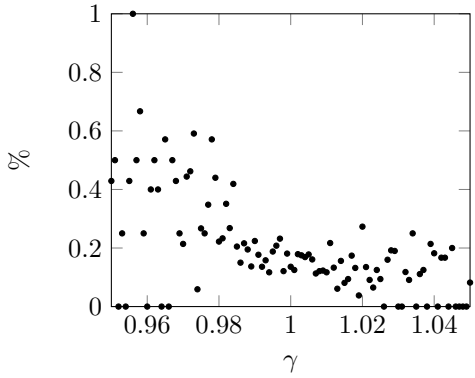

Second, to address the problem of size imbalance that is especially present in tails, we will transform the data and calculate a proportion of each yield that the algorithm predicted to each class. Results are plotted on Figs. \ref{fig:C0_predictions}, \ref{fig:C1_predictions}, and \ref{fig:C2_predictions}.

Dots show a proportion of each prediction class within the yield. For instance, value of $\text{C0}(\gamma=0.952)=1.0$ indicates that all predictions at that yield were classified into C0.{\footnote{In this concrete case, $1.0$ was a result of only 1 observation that was (falsely) predicted as C0.}} Since the data is given in relative terms, it is prone to outliers, especially if a number of observations for a yield was low. Irrespective of outliers, the plots clearly show that higher yields are, indeed, more heavily associated with C1 predictions and negative yields with C2 predictions.

\subsubsection{OLS Regression\label{sec:ols_regression}}
\noindent Let us test the relationship between yields and predicted class-proportions for each class with a simple linear regression model

\begin{equation*}
  y_{j,i} = \beta_{j,0} + \beta_{j,1} \cdot x_i + \epsilon_{j,i}.
\end{equation*}

Variable $x_i$ denotes a yield of observation $i$ and $y_{j,i}$ denotes a proportion of a yield that was classified to a class $j=(0,1,2)$. Models were estimated with OLS estimator. p-values are given in the brackets.

\begin{equation*}
    y_0 = \underset{(0.0158)}{1.9251} - \underset{(0.0577)}{1.4878}x \ \overline{R}^2=0.048
\end{equation*}

\begin{equation*}
    y_1 = -\underset{(0.0001)}{4.6905} + \underset{(0.0001)}{5.0495}x \ \overline{R}^2=0.493
\end{equation*}

\begin{equation*}
    y_2 = \underset{(0.0001)}{3.7654} - \underset{(0.0001)}{3.5617}x \ \overline{R}^2=0.344
\end{equation*}

Estimation results imply that there exists statistically significant linear relationship between yields and class-predictions for classes C1 and C2. Linear model turned out to be inadequate for predictions of C0. This means that farther in tails we go in both directions, more likely for the class prediction to be correct, that is either C1 or C2. As a corollary, exogenously given class-prediction probabilities would poorly capture the effect of different yields on outcomes. Consequently, a dependence of class-predictions on yields should be modelled more complex and separately for each class.

\subsubsection{Multinomial Logit}
\noindent To complement OLS regressions and complete the section, we ran a simple multinomial logit of the form

\begin{equation*}
  y_{j,i} = \beta_{j,0} + \beta_{j,1} \cdot x_{1,i,j} + \epsilon_{j,i}
\end{equation*}

with $y_i \in \{0=\text{C0},1=\text{C1},2=\text{C2}\}$ indicating a predicted class and $x_i$ is a yield of an observation $i=(1,...,N)$ and $j=(0,1,2)$. Results of the regression are shown in the Table \ref{tab:logit_regression}. C0 is taken as the benchmark alternative.

\begin{table}[h]
  \caption{Multinomial Logit Regression Summary}
  \label{tab:logit_regression}
  \centering
  \begin{tabular}{|l|c|c|c|}
      \hline
       & AVG & Prediction = C1 & Prediction = C2   \\
      \hline coef & & -22.8763 (0.0000) & 14.9750 (0.000)\\
      \hline yield & 1.003 & 22.4499 (0.0000) & -16.0069 (0.000)\\
      \hline LL & \multicolumn{3}{r|}{-3151.802} \\
      \hline LLR test: $\chi^2$ & \multicolumn{3}{r|}{214.596 (0.0000)} \\
      \hline
  \end{tabular}
\end{table}

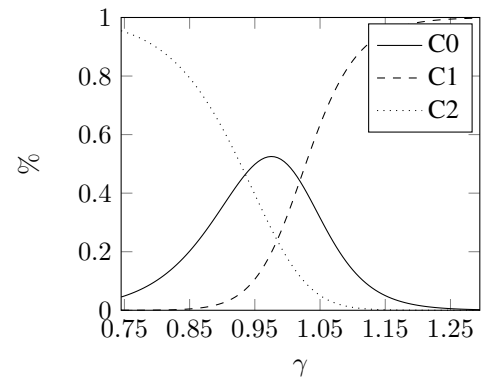
\begin{figure}[!h]
  \centering
  \pgfplotstableread[col sep=comma,]{res/class_probability_predictions.csv}\datatable
      \begin{tikzpicture}
          \begin{axis}[
              xmin=0.745, xmax=1.295, ymin=0, ymax=1,
              width=2.5in,
              xlabel=$\gamma$,
              ylabel=$\%$,
              xtick={0.75,0.85,...,1.29},scaled x ticks=false,
              ]
              \addplot [
                  color=black,
                  solid,
                  mark size=1pt
                  ] table[x=yield, y=0]{\datatable};
              \addplot [
                  color=black,
                  dashed,
                  mark size=1pt
                  ] table[x=yield, y=1]{\datatable};
              \addplot [
                  color=black,
                  dotted,
                  mark size=1pt
                  ] table[x=yield, y=2]{\datatable};
              \legend{C0,C1,C2}
          \end{axis}
      \end{tikzpicture}
  \caption{Class Probability Predictions}
  \label{fig:class_probability_predictions}
\end{figure}

Results are in line with OLS estimates and confirm that an increase in a yield is associated with a significant improvement of odds ratio in favor of C1 prediction in relation to the C0 and a significant deterioration of odds ratio of C2 prediction in relation to the base C0. This means that increasing the yield increases the probability of C1 prediction, while its decrease increases the probability of C2 prediction. To make the case more telling, we calculated probability predictions for all three classes on a wider domain and show results on the Fig. \ref{fig:class_probability_predictions}.

\subsection{Class Level Analysis}
\noindent Let’s check if the current findings translate to the level of particular types of predictions as well. Each yield $x_i$ is rounded with the factor $m$ to become $x_i=m \lceil \frac{x_i}{m} \rfloor$. Setting of $m$ is a trade-off between the round-off error that is caused with the rounding and a creation of meaningful groups for analysis. We consider $m=0.001$ or $0.1\%$ as a good compromise between the two.

Let $T(x_{i,j}^T\in X)$ and $P(x_{i,j}^P \in X)$ be two multisets of observed and predicted yields $x$ with $i=(1,...,N)$ and $j \in \{0,1,2\}$. $x_{i,j}^T =x_{i,j}^P$ indicates a correct prediction.

Let $C_{j}^T(\gamma)$ be a number of all observations from class $T_j$ with a yield equal to $\gamma$ for all classes $j = (0,1,2)$ and let $C_{j}^P(\gamma)$ be a number of all predictions from class $P_j$ with a yield equal to $\gamma$ for all classes $j = (0,1,2)$. Then $y_{j,k}(\gamma) = \frac{P_{j,k,\gamma}}{T_{j,\gamma}}$ with $k =(0,1,2)$ and $j = (0,1,2)$ is a proportion of $k$ class predictions of class $j$ per yield $\gamma$.

\begin{table*}[h]
    \centering
    \caption{Class Level Analysis}
    \label{Table:class_level_analysisII}
    \begin{tabular}[c]{|l|c|c|c|}
        \hline
        &P0 & P1 & P2\\
        \hline
        T0 
            & \begin{tabular}{@{}l@{}}$y=\underset{(0.7476)}{0.4079}+\underset{(0.9403)}{0.0948}x$ \\ $R^2=0.0002$\end{tabular} 
            & \begin{tabular}{@{}l@{}}$y=-\underset{(0.0095)}{2.4136}+\underset{(0.0036)}{2.7337}x$ \\ $R^2=0.2289$\end{tabular} 
            & \begin{tabular}{@{}l@{}}$y=\underset{(0.0383)}{3.0056}-\underset{(0.0511)}{2.8286}x$ \\ $R^2=0.1766$\end{tabular} \\
        \hline
        T1
            & \begin{tabular}{@{}l@{}}$y=\underset{(0.4453)}{2.2092}-\underset{(0.5180)}{1.8131}x$ \\ $R^2=0.0170$\end{tabular} 
            & \begin{tabular}{@{}l@{}}$y=-\underset{(0.2256)}{3.7526}+\underset{(0.1663)}{4.1783}x$ \\ $R^2=0.0656$\end{tabular} 
            & \begin{tabular}{@{}l@{}}$y=\underset{(0.0820)}{2.5434}-\underset{(0.0941)}{2.3652}x$ \\ $R^2=0.0675$\end{tabular} \\
        \hline
        T2
            & \begin{tabular}{@{}l@{}}$y=-\underset{(0.7459)}{1.5705}+\underset{(0.6748)}{2.0985}x$ \\ $R^2=0.0052$\end{tabular} 
            & \begin{tabular}{@{}l@{}}$y=\underset{(0.7348)}{1.4877}-\underset{(0.7666)}{1.3420}x$ \\ $R^2=0.0035$\end{tabular} 
            & \begin{tabular}{@{}l@{}}$y=\underset{(0.8056)}{1.0828}-\underset{(0.8676)}{0.7566}x$ \\ $R^2=0.0009$\end{tabular} \\
        \hline
    \end{tabular}
\end{table*}

Let’s test these relationships with a simple linear regression model of the form 

\begin{equation*}
    y_i = \beta_0 + \beta_1 \cdot x_i + \epsilon_i    
\end{equation*}

with variable $x_i$ denoting a yield and $y_i$ a corresponding $y_{j,k}(\gamma)$. Equations were estimated with the OLS estimator and results are shown in the Table \ref{Table:class_level_analysisII} with $\text{p-values}$ in the brackets.\footnote{In the table, T stands for a true class and P stands for a predicted class. A cryptic cell T0P1 then reads as a false C1 prediction of a true C0 class.}

Signs and magnitudes of regression coefficients are congruent with prior expectations. That is, a rise in a yield is associated with higher probability of the observation to become classified as C1 and significantly lower as C2. However, the models in general lack an appropriate statistical significance that restrain us from making a firm conclusion. A notable exceptions are T0P1 and T0P2 whose estimated coefficients are as expected and also highly statistically significant.

One possible explanation might be a small number of cases that makes an estimation more vulnerable to outliers. For instance, T2P1 has only 299 cases that are split in 3 groups and spread to 31 bins. This makes on average 3.25 elements per bin in a group. Given that observations are not split evenly on the domain, could, consequently, lead to significant dark spots, especially in tails.

\subsection{Majority without a Majority}
\noindent The algorithm makes classifications into groups based on the softmax function. Each observation is classified into a group such that $j^\ast \in \{1,...,J\} = \underset{k}{\text{arg max}} \ \text{Pr}(x_i \in X_k)$ for $k = (1,...,J)$. For a number of classes $J \geq 2$ the necessary condition for the best performing class is $\text{Pr}(j^\ast | x) \geq \underset{\delta \to 0}{\text{lim}} \left( \frac{1}{J}+ \delta \right)$ with $J$ equal to number of classes. In our case of 3 classes the minimal sufficient majority that led to the classification was $0.3377 \approx \frac{1}{3}+\delta$ for a true positive C0 case with a yield of 0.13\%.

Say that $\text{Pr}(j^\ast | x_i) \geq \alpha$ with $\alpha \in (0,1)$ is required for the classification of $x_i$ to $j^\ast$. Any value of $\alpha > \frac{1}{J}$ would induce a stronger requirement for classification. 

Table \ref{Table: Approval Rate} shows results for $\alpha = 0.95$. The change resulted in a significant boost of performance with an accuracy rate at 88.7\%. The precision of C1, the only class that triggers a trade, also improved and gained to 0.933 with only one falsely classified element that was from C0.\footnote{Besides, the yield of this element was also positive, though minimal at 0.1\%.} The recall score also improved to 0.875. The change of $\alpha$ made false predictions solely limited to adjacent classes.

\begin{table}[!htp]
  \centering
  \caption{95\% Approval Rate}
  \label{Table: Approval Rate}
    \begin{tabular}{|c|c|c|c|c|}
      \hline True / Prediction  & 0 & 1 & 2 & SUM\\
      \hline 0 & 46 & 1 & 5 & 52\\
      \hline 1 & 2 & 14 & 0 & 16\\
      \hline 2 & 0 & 0 & 3 & 3\\
      \hline SUM & 48 & 15 & 8 & 71\\
      \hline
    \end{tabular}
  \end{table}

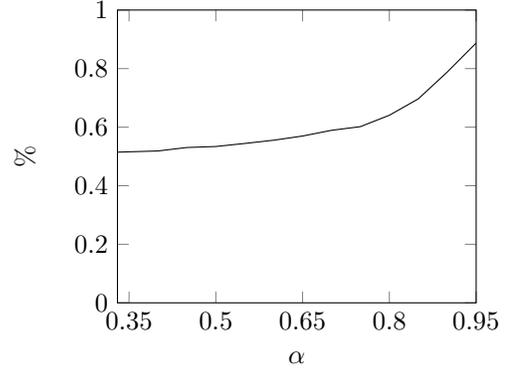
\begin{figure}[!htp]
    \centering
    \pgfplotstableread[col sep=comma,]{res/majority_no_majority.csv}\datatable
      \begin{tikzpicture}
        \begin{axis}[
            xmin=0.33, xmax=0.95, ymin=0, ymax=1,
            width=2.5in,
            xlabel=$\alpha$,
            ylabel=$\%$,
            xtick={0.35,0.5,...,0.9},scaled x ticks=false,
            ]
            \addplot [
                color=black,
                mark size=1pt
                ] table[x=alpha, y=correct]{\datatable};
        \end{axis}
    \end{tikzpicture}
    \caption{Correct Predictions per $\alpha$}
    \label{fig:proportion_of_correct_predictions}
\end{figure}

To see how the performance of the algorithm varies in relation to $\alpha$, we plotted a graph (Fig. \ref{fig:proportion_of_correct_predictions}) that shows a proportion of correct predictions given the $\alpha$. The curve is strictly increasing indicating that an increase in the required score increases a proportion of correct predictions, ceteris paribus.

\begin{figure}[!htp]
    \centering
    \pgfplotstableread[col sep=comma,]{res/majority_no_majority.csv}\datatable
    \begin{tikzpicture}
        \begin{axis}[
            xmin=0.33, xmax=0.95, ymin=0, ymax=0.4,
            width=2.5in,
            xlabel=$\alpha$,
            ylabel=$\%$,
            xtick={0.35,0.5,...,0.9},scaled x ticks=false,
            ]
            \addplot [
                color=black,
                mark size=1pt
                ] table[x=alpha, y=c1]{\datatable};
        \end{axis}
    \end{tikzpicture}
    \caption{C1 Predictions per $\alpha$}
    \label{fig:Proportions_of_predictions_c1}
\end{figure}
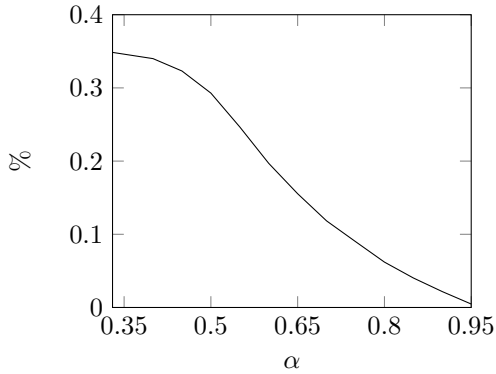

However, the gains came with a price. The most obvious is a significant drop in a number of observations that were classified. For $\alpha=0.95$, for instance, the number of classifications fell to about 2.2\% of the whole dataset (Fig. \ref{fig:Proportions_of_predictions_all}). That, consequently, led to a drop in C1 predictions as well (Fig. \ref{fig:Proportions_of_predictions_c1}).

\begin{figure}[!htp]
    \centering
    \pgfplotstableread[col sep=comma,]{res/majority_no_majority.csv}\datatable
    \begin{tikzpicture}
        \begin{axis}[
            xmin=0.33, xmax=0.95, ymin=0, ymax=1,
            width=2.5in,
            xlabel=$\alpha$,
            ylabel=$\%$,
            xtick={0.35,0.5,...,0.9},scaled x ticks=false,
            ]
            \addplot [
                color=black,
                mark size=1pt
                ] table[x=alpha, y=all]{\datatable};
        \end{axis}
    \end{tikzpicture}
    \caption{All Predictions per $\alpha$}
    \label{fig:Proportions_of_predictions_all}
\end{figure}
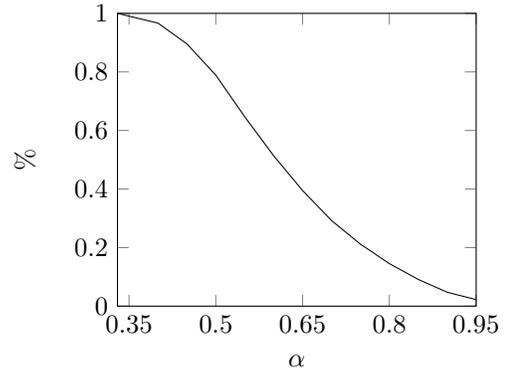

Both curves on graphs look strictly decreasing in $\alpha$ proving the negative relation between the $\alpha$ and number of predictions. Given both graphs we can say that increasing the requirement trades quantity for quality.

To derive a yield-maximization level of $\alpha$ assume that our classification problem can be simplified to two states $P \in \{0,1\}$ with $\gamma \in \{\overline{\gamma_0}, \overline{\gamma_1}\}$ denoting an average yield for each of the two states with $\overline{\gamma_1} \gg \overline{\gamma_0}$ and $\phi = (0,1)$ a proportion of C1 predictions in relation to all predictions. Then all predictions could be written as $P = \phi \cdot f(\alpha) + (1 - \phi) \cdot f(\alpha)$. Since only predictions of C1 trigger an activity, the second part on the right hand-side can be dropped since its expected value is 0. Assume further that $f(\alpha)$ is continuous, differentiable, non-negative and strictly monotonically decreasing function that relates number of predictions to $\alpha$ and that $g(\alpha)$ is monotonically increasing, continuous, non-negative and differentiable function that relates proportion of correct predictions to $\alpha \in \mathcal{R} = [0,1]$. Then the expected value can be written as $E(Y) = \phi \cdot f(\alpha)(g(\alpha) \cdot \gamma_{1} + (1-g(\alpha)) \cdot \gamma_0)$, from where follows
  
\begin{multline*}
    \frac{\partial E(Y,\alpha)}{\partial \alpha} = (\gamma_1 - \mu_0) (f'(\alpha) \cdot g(\alpha) + f(\alpha) \cdot g'(\alpha) )\\
    + \gamma_0 f'(\alpha) = 0
\end{multline*}

For the sake of simplicity say that $\gamma_0=0$, then the equation simplifies to $\frac{\partial E(Y,\alpha | \gamma_0=0)}{\partial \alpha} = \gamma_1(f'(\alpha) \cdot g(\alpha) + f(\alpha) \cdot g'(\alpha) ) = 0$. We know that $f'(\alpha) < 0$ and $g'(\alpha) > 0$, while $f(\alpha) \geq 0$ and $g(\alpha) \geq 0$. Setting $\gamma_0=0$ leads to $\gamma_1 \gg 0$. To solve the equation we set $f'(\alpha) \cdot g(\alpha) + f(\alpha) \cdot g'(\alpha) = 0$. The equation has a theoretical solution given the non-positive value of the first product and non-negative of the second.
  
In general it holds that $\frac{y'(x)}{y(x)} \approx \underset{\Delta x \rightarrow 0}{\text{lim}} \{\frac{\% \Delta f}{\% \Delta x} \cdot \frac{1}{x}\} \approx \underset{\Delta x \rightarrow 0}{\text{lim}} \{\frac{\% \Delta f}{\Delta x}\}$, given that $\Delta x = x \cdot \% \Delta x$.
  
In our case $f'(\alpha) \cdot g(\alpha) + f(\alpha) \cdot g'(\alpha) = \frac{f'(\alpha)}{f(\alpha)} + \frac{g'(\alpha)}{g(\alpha)} = 0$. The condition is satisfied at $\alpha^{\ast} = \alpha \in [0,1 ] \iff \underset{\Delta \alpha \rightarrow 0}{\text{lim}} \frac{\% \Delta f(\alpha) + \% \Delta g(\alpha)}{\Delta \alpha} = 0$. The written does not guarantee a unique solution.

\section{Trading with the Algorithm}
\noindent To test the functioning of the algorithm, we ran two simulation-based experiments that mimic trading in a real-time. Both experiments share the same setup. First, we collected information about a stock-price movement for an initial hour of trading. Second, the data was processed with the trained MNET-2 algorithm. Third, a trading decision was made based on the algorithm’s prediction. Unexploited opportunities expired and became valueless. Fourth, any initiated position was closed at the close of the trading or at the last possible trade with the corresponding asset. To simplify the computation, the assets were assumed to be infinitely divisible.

\subsection{One Investment at a Time}
\noindent In the first example, only one investment per time was allowed. Investment opportunities arrived sequentially, one-by-one, and all predicted buys were initiated.

Let $X^P = \{x_i(p^0_i,E(p^1_i)) | \frac{{E(p^1_i})} {p^0_i}\geq 1.02 \}$ for $i = 1,2,...,N$ with $p_i^0$ price of share $i$ at the open, $p_i^1$ a corresponding price at the close be a set of all predicted trades with $E(p)$ denoting an expected price change of an asset $i$. Let an initial investment equal $v_0$, the largest possible investment into a single position be limited to $\beta$ and $\{x^P\}_{t=1}^T$ be a sequence of all predicted trades. Then the investment pipeline can be written as

\begin{eqnarray*}
    v=\sum_{t=1}^{T} v_{{t-1}} + \frac{p(x^P_t)^1-p(x^P_t)^0} {p(x^P_t)^0}\\
    \times \max(0,\min(v_{t-1},\beta))
\end{eqnarray*}

with $p_t^0,p_t^1$ indicating stock prices at the open and the close of a position at the iteration $t=1,2,...,T$.

\subsection{Multiple Investments at a Time}
\noindent Limiting number of investments to one per time-interval is of little practical interest. In practice, at each time-period, like a day in our case, multiple trade opportunities exist and compete for financial resources.

Let $X^P = \{x_{i,t}(p^0_{i,t},E(p^1_{i,t})) | \frac{{E(p^1_{i,t}})} {p^0_{i,t}}\geq 1.02\}$ with $i = 1,2,...,N$ and $t = 1,...,T$ be a set of all predictions at each time $t$ and let multiple investments at $t$ be made in equal amounts $s$ that depend on an accumulated amount at time $v_t$ and a number of initiated investments. This is a reasonable assumption given our inability to rank yields. Assume further that there exists an upper limit $\beta$ that can be invested into a single asset. The upper limit is usually invoked to contain the risk. The invested amount into an asset at time $t$ can then be calculated as $s_t(v_t,|x_{\cdot,t}|\, | \beta) = \min \left(\beta, \frac{v_t}{|x_t|}\right)$. Since no shorting is allowed $s_t$ should be strictly non-negative $s_t =
  \begin{cases}
      s_t,&{\text{if}}\ s_t > 0,\\
      {0,}&{\text{otherwise}}
    \end{cases}$. Then the pipeline of trading can be written as

\begin{equation*}
      v = \sum_{t=1}^T \left[ v_{t-1} +  \sum_{j=1}^{J} \frac{p(x^P_{j,t})^1-p(x^P_{j,t})^0} {p(x^P_{j,t})^0} \cdot s_t \right].
\end{equation*}

\subsection{Simulations and Results}
\noindent We ran two sets of experiments, one for each architecture. Initial position of the first example was $v_0^1=1000\mathrm{USD}$, while in the second example initial investment was increased to $v_0^2=50,000\mathrm{USD}$. Simulations were ran on real data taken between January 14 and July 1, 2022. Altogether the dataset included 3179 observations that were realized within 114 trading days.

For the first architecture with investment opportunities arriving sequentially their order could have affected the result due to the accumulation of resources. To test for the effect we ran several repetitions with a shuffled order. However, it turned out that the effect was negligible in size. Time unit of the second simulation was (trading) day where trading opportunities arrived as they did on that real trading day. The order of opportunities in such configuration is irrelevant since all intended trades of a day are opened and closed at once. Results are given in the Table \ref{tab:simulation_results_trading}.

\begin{table*}[!htp]
  \caption{Simulation Results: Trading}
  \label{tab:simulation_results_trading}
  \centering
  \begin{tabular}{|c|c|c|c|c|c|c|}
      \hline
       Type & \multicolumn{3}{c|}{One Investment at Time} & \multicolumn{3}{c|}{Multiple Investments at Time}   \\
       \hline & Amount in US\$ & Trades & Amount per Trade in US\$ & Amount in US & Trades & Amount per Trade in US\$   \\
      \hline All & 10,465.88 & 3,179 & 3.29 & 58,831.79 & 3,179 & 18.51\\
      \hline Predicted C1 & 11,397.44 & 1,108 & 10.29 & 73,917.22 & 1,108 & 66.71\\
      \hline True C1 & 19,880.95 & 566 & 35.13 & 278,546.66 & 566 & 492.13\\
      \hline Random: 50\% & 5,710.38 & 1,583 & 3.61 & 57,961.38 & 1,565 & 37.04\\
      \hline Random: 50\% & 4,802.04 & 1,608 & 2.99 & 58,443.35 & 1,565 & 37.34\\
      \hline Random: 33\% & 3,695.26 & 1,028 & 3.59 & 58,589.26 & 1,019 & 57.50\\
      \hline Random: 33\% & 4,238.29 & 1,058 & 4.01 & 56,755.77 & 1,019 & 55.70\\
      \hline Random: 33\% & 4,008.80 & 1,007 & 3.98 & 56,240.11 & 1,019 & 55.19\\            
      \hline
  \end{tabular}
\end{table*}

“All” is a case that initiated all possible trades. “Predicted C1” is a case that initiated trades that were predicted as C1 by the algorithm. “True C1” is a case that initiated trades that yielded at least 2\%. The decision for trading an asset in the remaining 5 simulation-runs was done by random with a probability of each trade as in the table.

Trading with the algorithm over the analyzed period resulted in a positive yield. Each investment of the first scenario had an upper limit of $\beta=1000\mathrm{USD}$ to prevent the system from exploding. In the first scenario, trading with the algorithm managed to turn initial 1000USD to nearly 11,400USD in 1108 trades for an average yield per trade of 0.938\%. This is close to the simulated average yield of 0.95\%. The algorithm outperformed all other simulation examples in absolute value and relative terms except a theoretical maximum. For instance, if going all trades long would on average yield 0.3\% per trade, while going long on True C1 would on average yield 3.34\% per transaction which is a clear top performing configuration.

In the second scenario 50,000USD was initially invested and the algorithm realized an aggregate yield of 47.8\%. It, again, executed 1108 trades but now in 114 trading days for 9.7 trades per a trading day. The average yield on an initial investment per day stood at 0.343\%. If going long in all transactions would yield “only” 17.7\% on aggregate or 0.143\% per day which is evidently inferior to the algorithm.

Trading with the algorithm significantly reduced number of transactions. Altogether, only 34.85\% of all possible trades were initiated. The practical consequence of a smaller number of trades are numerous: (1) it shortens execution time, (2) allows much larger investment per trade for a given capital and (3) much smaller capital requirement for a given size of an investment, (4) enables focusing on high-yield opportunities, (5) makes portfolio smaller and more manageable.

First, it is self-evident that an execution time for some fixed number of trades is reverse to the number of trades that are done at once, ceteris paribus. In our case, 1108 days would be needed if only one trade was done in a day against 114 if allowing multiple trades at once or 1108 days if investing according to the algorithm against 3179 if going all positions long. Second, if an investor has some fixed amount of capital allocated for investing, which is usually the case, then smaller number of trades would by definition lead to larger investments per trade. Given that investing is not a subject of commutativity between a number of trades and an investment per trade, then making larger investments in smaller number of assets with higher yields could lead to higher aggregate return. Third, if a request is for each position to be of some fixed size, then having invested in a smaller number of assets would by definition require smaller amount of capital that is needed to initiate all positions. This is desirable feature since capital comes with costs. Fourth, smaller number of investments makes it possible to increase a return by passing on lower-yield (though still positive) opportunities and making larger investments in high-yield assets. In fact, this is the core of the paper. Fifth, it should not come as a surprise that a smaller portfolio of investments is more manageable. Although, the technological innovation may diminish the size of the effect.

\section{Conclusion}
\noindent In the paper the asset-trading was considered as an image classification problem and was processed with the CNN-based MobileNet-V2 neural network. Items were classified in three classes that marked a clear drop in price, a clear rise in price or too volatile to call based on the price change. We did not aim to model the level of the price change. The model was trained and tested on the largest NASDAQ-listed stocks by market capitalization using intraday data at a 5-minute time interval.

Trading with the algorithm outperformed the market during the tested period as well as all trading configurations of the testing with an exception of the theoretical maximum. The algorithm was able to distinguish well between the two opposing classes and allocated the majority of false classifications into the third, yield-neutral class. Using the algorithm significantly reduced number of transactions to roughly one-third. All testing was done on an in-sample dataset.

During the training, the best performing model of the validation phase achieved 49.36\% accuracy for three balanced classes in size. The accuracy of the best performing model during the testing stood at 51.5\%. We showed that the score could be significantly improved if placing a stricter conditions for classification. However, that led to a sharp drop in a number of classifications themselves. It was clearly a case of a trade-off between the quality and quantity.

The principal result of the paper is that there exists a high-level relationship between the first hour of trading and the close that the CNN-based MobileNet-V2 model managed to extract. Given results a reasonable conclusion would be that the algorithm is useful as long as the dataset is distributed as closely as possible to the data that the algorithm is expected to predict.


\bibliography{references}

\begin{thebibliography}{10}
\providecommand{\url}[1]{#1}
\csname url@samestyle\endcsname
\providecommand{\newblock}{\relax}
\providecommand{\bibinfo}[2]{#2}
\providecommand{\BIBentrySTDinterwordspacing}{\spaceskip=0pt\relax}
\providecommand{\BIBentryALTinterwordstretchfactor}{4}
\providecommand{\BIBentryALTinterwordspacing}{\spaceskip=\fontdimen2\font plus
\BIBentryALTinterwordstretchfactor\fontdimen3\font minus
  \fontdimen4\font\relax}
\providecommand{\BIBforeignlanguage}[2]{{%
\expandafter\ifx\csname l@#1\endcsname\relax
\typeout{** WARNING: IEEEtran.bst: No hyphenation pattern has been}%
\typeout{** loaded for the language `#1'. Using the pattern for}%
\typeout{** the default language instead.}%
\else
\language=\csname l@#1\endcsname
\fi
#2}}
\providecommand{\BIBdecl}{\relax}
\BIBdecl

\bibitem{SubinReinicke_2022}
\BIBentryALTinterwordspacing
S.~Subin and C.~Reinicke, ``Dow closes 500 points lower after the fed delivers
  another aggressive rate hike,'' CNBC, Sep. 20, 2022 [Online]. [Online].
  Available:
  \url{https://www.cnbc.com/2022/09/20/stock-market-futures-open-to-close-newshtml.html}
\BIBentrySTDinterwordspacing

\bibitem{bodie1976common}
Z.~Bodie, ``Common stocks as a hedge against inflation,'' \emph{The journal of
  finance}, vol.~31, no.~2, pp. 459--470, 1976.

\bibitem{fama1977asset}
E.~F. Fama and G.~W. Schwert, ``Asset returns and inflation,'' \emph{Journal of
  financial economics}, vol.~5, no.~2, pp. 115--146, 1977.

\bibitem{campbell1987stock}
J.~Y. Campbell, ``Stock returns and the term structure,'' \emph{Journal of
  financial economics}, vol.~18, no.~2, pp. 373--399, 1987.

\bibitem{keim1986predicting}
D.~B. Keim and R.~F. Stambaugh, ``Predicting returns in the stock and bond
  markets,'' \emph{Journal of financial Economics}, vol.~17, no.~2, pp.
  357--390, 1986.

\bibitem{french1987expected}
K.~R. French, G.~W. Schwert, and R.~F. Stambaugh, ``Expected stock returns and
  volatility,'' \emph{Journal of financial Economics}, vol.~19, no.~1, pp.
  3--29, 1987.

\bibitem{barberis2000investing}
N.~Barberis, ``Investing for the long run when returns are predictable,''
  \emph{The Journal of Finance}, vol.~55, no.~1, pp. 225--264, 2000.

\bibitem{stambaugh1999predictive}
R.~F. Stambaugh, ``Predictive regressions,'' \emph{Journal of financial
  economics}, vol.~54, no.~3, pp. 375--421, 1999.

\bibitem{welch2008comprehensive}
I.~Welch and A.~Goyal, ``A comprehensive look at the empirical performance of
  equity premium prediction,'' \emph{The Review of Financial Studies}, vol.~21,
  no.~4, pp. 1455--1508, 2008.

\bibitem{barberis2003survey}
N.~Barberis and R.~Thaler, ``A survey of behavioral finance,'' \emph{Handbook
  of the Economics of Finance}, vol.~1, pp. 1053--1128, 2003.

\bibitem{eapen2019novel}
J.~Eapen, D.~Bein, and A.~Verma, ``Novel deep learning model with cnn and
  bi-directional lstm for improved stock market index prediction,'' in
  \emph{2019 IEEE 9th annual computing and communication workshop and
  conference (CCWC)}.\hskip 1em plus 0.5em minus 0.4em\relax IEEE, 2019, pp.
  0264--0270.

\bibitem{tsantekidis2017forecasting}
A.~Tsantekidis, N.~Passalis, A.~Tefas, J.~Kanniainen, M.~Gabbouj, and
  A.~Iosifidis, ``Forecasting stock prices from the limit order book using
  convolutional neural networks,'' in \emph{2017 IEEE 19th conference on
  business informatics (CBI)}, vol.~1.\hskip 1em plus 0.5em minus 0.4em\relax
  IEEE, 2017, pp. 7--12.

\bibitem{sim2019}
\BIBentryALTinterwordspacing
H.~S. Sim, H.~I. Kim, and J.~J. Ahn, ``Is deep learning for image recognition
  applicable to stock market prediction?'' \emph{Complexity}, vol. 2019, p.
  4324878, 2019. [Online]. Available:
  \url{https://doi.org/10.1155/2019/4324878}
\BIBentrySTDinterwordspacing

\bibitem{gudelek2017deep}
M.~U. Gudelek, S.~A. Boluk, and A.~M. Ozbayoglu, ``A deep learning based stock
  trading model with 2-d cnn trend detection,'' in \emph{2017 IEEE symposium
  series on computational intelligence (SSCI)}.\hskip 1em plus 0.5em minus
  0.4em\relax IEEE, 2017, pp. 1--8.

\bibitem{sezer2018algorithmic}
O.~B. Sezer and A.~M. Ozbayoglu, ``Algorithmic financial trading with deep
  convolutional neural networks: Time series to image conversion approach,''
  \emph{Applied Soft Computing}, vol.~70, pp. 525--538, 2018.

\bibitem{long2019deep}
W.~Long, Z.~Lu, and L.~Cui, ``Deep learning-based feature engineering for stock
  price movement prediction,'' \emph{Knowledge-Based Systems}, vol. 164, pp.
  163--173, 2019.

\bibitem{chen2021novel}
W.~Chen, M.~Jiang, W.-G. Zhang, and Z.~Chen, ``A novel graph convolutional
  feature based convolutional neural network for stock trend prediction,''
  \emph{Information Sciences}, vol. 556, pp. 67--94, 2021.

\bibitem{hu2021survey}
Z.~Hu, Y.~Zhao, and M.~Khushi, ``A survey of forex and stock price prediction
  using deep learning,'' \emph{Applied System Innovation}, vol.~4, no.~1, p.~9,
  2021.

\bibitem{sezer2020financial}
O.~B. Sezer, M.~U. Gudelek, and A.~M. Ozbayoglu, ``Financial time series
  forecasting with deep learning: A systematic literature review: 2005--2019,''
  \emph{Applied soft computing}, vol.~90, p. 106181, 2020.

\bibitem{sandler2018mobilenetv2}
M.~Sandler, A.~Howard, M.~Zhu, A.~Zhmoginov, and L.-C. Chen, ``Mobilenetv2:
  Inverted residuals and linear bottlenecks,'' in \emph{Proceedings of the IEEE
  conference on computer vision and pattern recognition}, 2018, pp. 4510--4520.

\bibitem{goodfellow2016deep}
I.~Goodfellow, Y.~Bengio, and A.~Courville, \emph{Deep learning}.\hskip 1em
  plus 0.5em minus 0.4em\relax MIT press, 2016.

\bibitem{yfinance_2022}
\BIBentryALTinterwordspacing
R.~Aroussi, \emph{yfinance}, ver. 0.1.85, 2022 [Online]. [Online]. Available:
  \url{https://github.com/ranaroussi/yfinance.git}
\BIBentrySTDinterwordspacing

\bibitem{mplfinance_2022}
\BIBentryALTinterwordspacing
D.~Goldfarb, \emph{mplfinance}, ver. 0.12.9b5, 2022 [Online]. [Online].
  Available: \url{https://pypi.org/project/mplfinance/}
\BIBentrySTDinterwordspacing

\bibitem{pmlr-v119-rice20a}
\BIBentryALTinterwordspacing
L.~Rice, E.~Wong, and Z.~Kolter, ``Overfitting in adversarially robust deep
  learning,'' in \emph{Proceedings of the 37th International Conference on
  Machine Learning}, ser. Proceedings of Machine Learning Research, H.~D. III
  and A.~Singh, Eds., vol. 119.\hskip 1em plus 0.5em minus 0.4em\relax PMLR,
  13--18 Jul 2020, pp. 8093--8104. [Online]. Available:
  \url{https://proceedings.mlr.press/v119/rice20a.html}
\BIBentrySTDinterwordspacing

\bibitem{JMLR:v15:srivastava14a}
\BIBentryALTinterwordspacing
N.~Srivastava, G.~Hinton, A.~Krizhevsky, I.~Sutskever, and R.~Salakhutdinov,
  ``Dropout: A simple way to prevent neural networks from overfitting,''
  \emph{Journal of Machine Learning Research}, vol.~15, no.~56, pp. 1929--1958,
  2014. [Online]. Available:
  \url{http://jmlr.org/papers/v15/srivastava14a.html}
\BIBentrySTDinterwordspacing

\bibitem{warde2013}
\BIBentryALTinterwordspacing
D.~Warde-Farley, I.~J. Goodfellow, A.~Courville, and Y.~Bengio, ``An empirical
  analysis of dropout in piecewise linear networks,'' 2013. [Online].
  Available: \url{https://arxiv.org/abs/1312.6197}
\BIBentrySTDinterwordspacing

\end{thebibliography}

\begin{IEEEbiography}[{\includegraphics[width=1in,height=1.25in,clip,keepaspectratio]{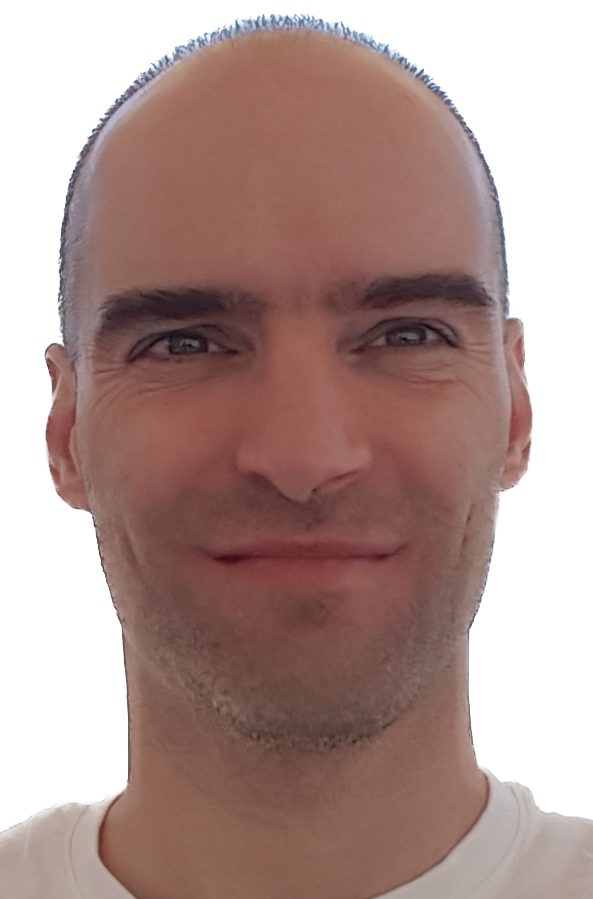}}]{Matej Steinbacher}
graduated in economics from the University of Maribor, Slovenia and continued with the post-graduate studies at the Faculty of Economics of the University of Ljubljana, Slovenia. His research is focused on studying network models in economics and finance.
\end{IEEEbiography}

\end{document}